# Response to "Comment on 'Zero and negative energy dissipation at information-theoretic erasure'"

Laszlo Bela Kish, Claes-Göran Granqvist, Sunil P. Khatri, Ferdinand Peper

**Abstract** We prove that statistical information theoretic quantities, such as information entropy, cannot generally be interrelated with the lower limit of energy dissipation during information erasure. We also point out that, in deterministic and error-free computers, the information entropy of memories does not change during erasure because its value is always zero. On the other hand, for information-theoretic erasure—*i.e*., "thermalization" / randomization of the memory—the originally zero information entropy (with deterministic data in the memory) changes after erasure to its maximum value, 1 bit / memory bit, while the energy dissipation is still positive, even at parameters for which the thermodynamic entropy within the memory cell does not change. Information entropy does not convert to thermodynamic entropy and to the related energy dissipation; they are quantities of different physical nature. Possible specific observations (if any) indicating convertibility are at most fortuitous and due to the disregard of additional processes that are present.

L.B. Kish (✉) and S.P. Khatri
Department of Electrical and Computer Engineering, Texas A&M University, College Station, TX 77843-3128, USA
e-mail: laszlo.kish@ece.tamu.edu, sunilkhatri@tamu.edu

C.G. Granqvist
Department of Engineering Sciences, The Ångström Laboratory, Uppsala University, P.O. Box 534, SE-75121 Uppsala, Sweden
e-mail: claes-goran.granqvist@angstrom.uu.se

F. Peper
CiNet, NICT, and Osaka University, 1-4 Yamadaoka, Suita, Osaka, 565-0871, Japan
e-mail: peper@nict.go.jp

## 1 Introduction

In a recent paper [1,2], we introduced "Information-Theoretic Erasure" with various device concepts and applied these schemes to show that Landauer's Principle or erasure dissipation is invalid. In the expanded version [1], we briefly addressed paper [3]. Concerning this issue (paper [3]) a critical Comment [4] was recently written.

In this Response, after first defining some essential terms and briefly summarizing the relevant results in our earlier article [1], we address the points where we agree with Comment [4] as well as the issues where we disagree with [3,4].

## 2 Information-Theoretic Erasure and Landauer's Principle

*2.1 Types of erasure of data in memories*

(a) *Secure erasure by resetting the bits to zero*. This is the type of erasure assumed in the original version of Landauer's Principle [3]; see Equation 2 below. This type of erasure is used only for security applications in computers because it is extremely slow and very energy-guzzling.

(b) "*Erasure" by writing-over* [5]. Here the memory bits are not reset; instead the blocks of the memory to be erased are designated as "free" and otherwise left alone, to be written over by new data that needs to be written. The number of address bits and the "erasure"-related dissipation scales as $\log_2 N$, where $N$ is the size of the whole memory. This type of "erasure" is used in computers; it is the fastest and requires minimal energy dissipation. The logarithmic scaling is in direct contradiction with Landauer's principle, see below.

(c) *Information-theoretic erasure* (ITE) [1,2]. This erasure is a randomization/thermalization process

with minimum energy dissipation, which can be as low as zero when erasure is done in a passive way; see the double-well example below. For ITE, bit errors are generated by thermal noise, which results in 50% chance for the values 0 and 1 after erasure and no information about the original memory content. On the other hand, it is well-known that Shannon's information entropy $S_\mathrm{I}$ now attains its *absolute maximum* and is given, for the case of $N$ bits, by

$$S_\mathrm{I} = \sum_{j=1}^{N}\sum_{m=0}^{1} p_{j,m} \ln\left(\frac{1}{p_{j,m}}\right) = \sum_{j=1}^{N}\sum_{m=0}^{1} 0.5 \ln\left(\frac{1}{0.5}\right) = N \quad (1)$$

so that the information entropy during ITE *can only increase or remain constant*. Here $p_{j,m}$ stands for the probability of the $j$-th bit being in the $m$-bit value. To check that this erasure principle is both physical and physically realizable, we introduced and analyzed two device concepts with regard to ITE: one with double-potential wells and another with capacitors [1,2]. We found that, when the working conditions of such a capacitor-based memory involved less than $k_\mathrm{B}T/2$ stored energy (and an equivalent negative thermal entropy $S_\mathrm{th}$) for holding the originally stored information, ITE (which is thermalization) entailed that the system of capacitors absorbed heat from the environment thus yielding negative energy dissipation. However, we emphasize that the energy dissipation of the shown ITE schemes is always positive when the control of the switches [6,7] arranging the erasure is also accounted for.

*2.2 Landauer's Principle*

The "classical" version of Landauer's Principle (quoted also in [3]) asserts that

$$\Delta Q_\mathrm{th} = T\Delta S_\mathrm{th} \geq -k_\mathrm{B}T\ln(2)\Delta S_\mathrm{I}, \quad (2)$$

where $\Delta Q_\mathrm{th}$ and $\Delta S_\mathrm{th}$ are produced heat and thermodynamic entropy, and $\Delta S_\mathrm{I}$ is the change of $S_\mathrm{I}$ during erasure. Equation 2 states that the change of information entropy can be "converted" into the lower limit of energy dissipation during the erasure of a memory. It should be emphasized that the greater-than-or-equal-to sign—rather than a greater-than sign—is very important because the equality must represent a physical possibility, at least at the conceptual level. One of the most straight-forward and relevant objections [5–7] to Equation 2 is that, in the case of equality, the memory cell's error probability is 50% even in the short-time limit, *i.e.*, the memory does not function and it is practically useless for the case of even larger energy dissipation; the limit of equality is unphysical in Equation 2.

ITE and Equation 1 are in direct contradiction with Landauer's Principle (Equation 2) because, even in the absence of any available negative thermal entropy $S_\mathrm{th}$, the minimum energy dissipation would be allowed to be negative during erasure, which obviously is not correct. This impossibility is illustrated by the example of passive ITE for double-well-potential based memories below [1,2]. A single memory cell is shown in Figure 1.

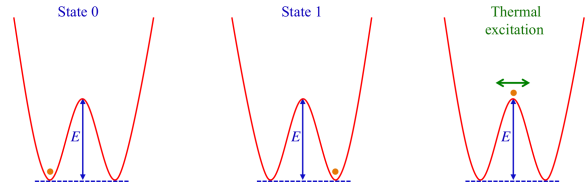

**Fig. 1** Passive information-theoretic erasure in a zero-energy-dissipation fashion by passively waiting during times that are much longer than the thermalization time constant at ambient temperature [1,2].

For the sake of simplicity, suppose that originally all bits are in the 1 state with $p_{j,1}=1$ and $p_{j,0}=0$, which means $S_\mathrm{I}=0$. Now, let us wait for a time $t_\mathrm{w} \gg \tau_0 \exp(E/k_\mathrm{B}T)$ until the double-wells are "thermalized" and $p_{j,1}=p_{j,0}=0.5$, implying $S_\mathrm{I}=N$ (bits) so that the information entropy of the memory has increased to $\Delta S_\mathrm{I}=N$ without any energy dissipation or energy investments or control. Landauer's principle (Equation 2) does not specify any restriction on the duration of erasure, and hence it applies here and yields that

$$\Delta Q_\mathrm{th} = T\Delta S_\mathrm{th} \geq -k_\mathrm{B}T\ln(2)\Delta S_\mathrm{I} = \\ = -Nk_\mathrm{B}T\ln(2) \gg 0, \quad (3)$$

which is incorrect because the energy dissipation during this erasure is always exactly zero.

*The examples above indicate how important is to check any proposed general mathematical principle in physics with thought-experiments and physical conceptual models embracing all of the essential details. Mathematics is infinitely richer than physics, and ultimately the Laws of Physics will select those few mathematical principles, models and solutions which are physical, that is, realistic. It is obvious from the considerations above that Landauer's mathematical principle (Equation 2) is unphysical.*

It should be noted that Landauer's Theorem has been criticized and refuted in many independent works and with different arguments [5–14]—including in well-known debates such as those involving Porod *et al.* [8–10]—and recently by Norton [11,12] in extensive studies.

In the next sections, we reflect on some of the most important comments in [4] and identify the major points where we agree and disagree with them and/or with his related work [3], and similar approaches by others, where information change is claimed to contribute to energy dissipation during erasure.

## 3 Points where we agree with [4]

In [3] the Landauer Principle (Equation 2) was expanded and generalized in an original way by introducing other types of entropy contributions in addition to the information entropy during erasure. They can contribute to the energy dissipation with zero, positive or negative values. Moreover [3] used a quantum system to store and erase classical information and, instead of Shannon entropy, used another statistical information measure. We agree with the Comments [4] concerning two points:

(a) Stimulated by the referee's comments on our work [2], we introduced the ice-cube-memory model [1], which proved that the early version (Equation 2) of Landauer's Principle is invalid. Bit value 1 was ice-phase, and bit value 0 was water-phase. Instead of ITE, we simply supposed a reset-to-zero operation by allowing the environment to melt the ice, which emerges as negative energy dissipation. While this simple trick proved the old Landauer Principle incorrect, we unfortunately misread paper [3] wherein the additional entropies in the model could account for the negative thermodynamic entropy and the related energy flow represented by the ice. Thus while we proved the usual (i.e. former) interpretation of the Landauer Principle (2) invalid, our model was fully in accordance with the extended principle [3].

(b) The other point of accord is the fact that, in a footnote in [3] the possibility of erasing memories by heating them and allowing them to thermalize is mentioned. This is a special case of ITE, which we regrettably failed to acknowledge. In our papers [1,2], we introduce the generalized ITE concept, without heating, and introduced active and passive ITE schemes with related device concepts. However, [3] was first to mention erasure by thermal randomization.

## 4 Points where we disagree with [3,4]

(a) Sections 2.1(b) and 2.2 above delineated two major erasure scenarios where we disagree with [3] and we believe that both versions of Landauer's Theorems—*i.e.*, Equation 2 and the expanded version [3]—do not work. In these simple classical physical systems, there is no other entropy source of compensation to account to restore Landauer principle in the fashion [3] is doing. The fact that [3] is using a quantum system does not make these consideration irrelevant because [3] works with classical information.

(b) One of the reasons for the dichotomous opinions is our assertion that statistical information measures are insufficient to describe dissipation in memories during erasure. A simple example is given below for erasure by resetting the bits to zero.

The erasure of a classical physical memory cannot in all cases depend on whether we know the data in the memory or not. Suppose first we know the data bits. This means that the Shannon information entropy or any other information measure is zero even if the memory is full of data because the probabilities of the bit being 0 or 1 bit in each memory cell are either 0 or 1, and this leads to zero information entropy (e.g. Equation 1). However, even after erasure, the system would be in a *known* deterministic state with all bits having now zero value, still leading to zero information entropy. Thus there was no information entropy loss, because there was no information to begin with. This fact highlights that statistical information measures are irrelevant.

(c) When we do not know the data in the memory, perhaps the deepest reason for not using information entropy to describe memory dissipation is Alfred Renyi's arguments about deterministic systems such as error-free computers [15]. The information entropy of deterministically generated data is less than or equal to the information entropy of the given algorithm and its initialization parameters. For example, let us suppose we generate $\pi$ with a simple deterministic algorithm and record its bits into a memory. All elements of the algorithm generating $\pi$ are known and deterministic, therefore the information entropy of the algorithm is zero! Even when we enter more and more bits of $\pi$ into the memory and the data size $N$ and corresponding erasure dissipation approaches infinity, the information entropy of this infinitely large random data sequence will be $\log_2 N$, the address required to identify the last digit. While the erasure dissipation scales with $N$, the information before erasure scales logarithmically. No physical mechanism exists to compensate for this non-existent, logarithmically scaling dissipation. *Reductio ad absurdum*.

## 5 Conclusions

Our main conclusion is that statistical information measures are irrelevant for treating the energy dissipation during memory erasure. This is an implication relevant for not only [3] but also for all the other papers in the literature that are assuming the validity of Landauer's principle.

Note, this fact does not prove that the Neumann entropy used in [3] is unable to treat the energy dissipation but it indicates that statistical information measures are irrelevant for the erasure dissipation. We showed contra-examples proving that no *general principle* interrelating information and energy dissipation can physically be justified.

We reiterate that it is unavoidable to test heuristic mathematical principles in physics by creating a concrete device or system concept for which the principle can be challenged, because a virtually infinite number of different mathematically valid models and principles can be created that are actually unphysical. Throughout our criticism of the Landauer Theorem and related issues, we followed this way of checking.

It is important to note that the mentioned conceptual scheme must reflect on *all* the essential aspect of the physical model. For example a "massless trapdoor", a "frictionless system" or a "thermal-noise-free device", that are assumptions which explicitly or implicitly are present in many high-profile scientific articles of today, are false assumptions, as are ignoring the energy requirement of control steps in a memory, Maxwell demon or Szilard engine [5,6].